\documentclass[preprint,12pt, authoryear]{mn2e}
\usepackage{graphicx}
\usepackage{amssymb}
\usepackage{footnote}
\usepackage{threeparttable}
\usepackage{epsfig}
\usepackage[usenames,dvipsnames,svgnames]{xcolor}
\usepackage{pslatex}
\usepackage{amsmath}
\usepackage{epstopdf}
\usepackage{multirow}
\usepackage{array}

\title[Novae grid model]{A photoionization model grid for novae: estimation of physical parameters}
\author[Mondal et al.]{Anindita Mondal$^{1}$,\thanks{E-mail: anindita12@bose.res.in} Ramkrishna Das$^{1}$, Gargi Shaw$^{2}$, Soumen Mondal$^{1}$\\
$^{1}$S N Bose National Centre for Basic Sciences, Salt Lake, Kolkata 700 106, India\\
$^{2}$Tata Institute of Fundamental Research, Mumbai 400 005, India}
\date{June 2018}

\pagerange{\pageref{firstpage}--\pageref{lastpage}} \pubyear{2018}

\begin{document}

\label{firstpage}

\maketitle
\begin{abstract}
We present here a method to estimate physical parameters of novae systems using an extensive grid of photoionization models for novae.
We use the photoionization code CLOUDY to construct grid of models covering a wide
range of different parameters, e.g. total hydrogen density ($n_H$), source temperature ($T_{BB}$) and luminosity ($L$),
inner radius ($R_{in}$) and thickness of ejecta ($\Delta R$), keeping other elements at solar metallicity. In this way, a total of 1792 models have been generated. From the model generated spectra which cover a wide wavelength region from ultra-violet to infrared, we calculate ratios of hydrogen and helium emission lines fluxes which are generally strong in novae spectra. We show that physical parameters associated with novae system could be estimated by comparing these line ratios with those obtained from observed spectra. We elaborate the idea with examples and estimate the parameter values in case of few other novae. The results of the grid model are available online.

\end{abstract}


\section{Introduction}
It is well known that novae refer to a close interacting binary system. The primary component
is a compact white dwarf (WD) which could be either a CO type or an ONe type and the secondary component is generally a main sequence
star or a late-type giant. This system is very close, with orbital periods $<$ 16 hr, allowing mass transfer from the
secondary star onto its companion. As a result, hydrogen-rich matter from the secondary is accreted onto the WD surface via an accretion disc.
Over a period of time, the accreted layer on the WD grows in mass, consequently pressure and temperature
at the base of the accreted layer rise gradually. When critical temperature and pressure are reached, thermonuclear burning of hydrogen set in which soon builds up to a thermonuclear runaway (TNR)
reaction releasing huge amount of energy ($\sim 10^{45}$ erg) in a very short period of time. This is commonly known
as nova outburst. The explosion is accompanied by ejection of matter with velocities of the order of few hundreds to thousands km/s in the form of discrete shell(s),
an optically thick wind, or as a combination of both. The detailed theory and development of nova outburst have been described in several
articles, e.g. Bode $\&$ Evans (2008); Starrfield et al. (2008); Warner (1995); BASI (2012).\\
\\
Observationally, the outburst is accompanied by a sudden rise in optical brightness,
generally with an amplitude of $\sim$ 7 to 15 magnitudes in 1-2 days. The peak luminosities may be as high as 10$^{4}$ - 10$^{5}$ $L_\odot$
above the quiescence phase brightness of the object. This is followed by a gradual decline in the light curve on time scales of
months to years. All aspects of the outburst are also manifested at various stages in the evolution of
nova spectra. At the early stage (fireball phase), when the ionization levels are low, the spectrum is generally
dominated by permitted, recombination lines of H I, He I, C I, O I, Fe I and N I. As the ejecta expands with time, density
of the ejecta decreases and layers closer to the central ionizing source are revealed and degrees of excitation and
ionization increase with time. Forbidden and high ionization emission lines are seen at this stage.
For example, prominent lines of [Fe VI], [Fe VII], [Ca V], [Mn XIV] and [Si VI] are seen in coronal phase, whereas, lines
of [Ne III], [O I], [Fe X], [Fe XIV], [Ca XV], [Ni XII] etc. are observed in nebular phase. As the nova approaches
its post-outburst quiescence phase, the ionization levels decrease once again.
Thus, study of observational properties of emitted spectra may help to understand the properties of the system.\\
\\
In few previous studies, attempts have been taken to explain observational properties such as light curves, characteristic times, formation
of spectrum in novae (e.g. Prialnik $\&$ Kovetz 1995; Yaron et al. 2005; Shara et al. 2010; Hachisu $\&$ Kato 2006).
Hauschildt et al. (1995, 1996, 1997) generated synthetic spectra using PHOENIX code to study the formation of spectral lines in novae. PHOENIX is a stellar atmosphere code that is designed to produce model spectra of stars for different effective temperatures, masses, metallicities, etc. Similarly, CLOUDY (Ferland et al. 2013, 2017) is an astrophysical plasma code that uses a given density and an input SED (e.g. stars, active galactic nuclei, etc.) to calculate various physical processes and predict the spectrum as the radiation interacts with gas and dust of known composition and geometry under a broad range of conditions.
Both the codes CLOUDY and PHOENIX, can be used to simulate 1D and 3D non-LTE models of novae spectra.
PHOENIX is basically used as a stellar atmosphere code, whereas, the photo-ionization code CLOUDY has a wider range of applicability with a larger atomic, molecular and chemical database. Therefore, CLOUDY has a potential to provide better results for novae system, where photoionization plays the most important role in generating the emission line spectra.\\
\\
In this paper, we calculate simple grid models of novae using the photoionization code CLOUDY (version c17.00 rc1) (Ferland et al., 2017).
Our aim is to investigate how the spectral emission line intensities change under different physical conditions and if the line ratios can be used to estimate the physical parameters from the observed spectra of novae. In order to do so, we construct a grid of 1792 models for different set of parameters associated with novae systems, viz. inner radius ($R_{in}$) of the ejected shell, thickness of ejected shell ($\Delta R$),
source temperature ($T_{BB}$), source luminosity ($L$) $\&$ H-density ($n_H$).
For each set of parameters, spectra are generated over a wide range of wavelength, from ultraviolet to infrared.
Next, we calculate the ratios of hydrogen and helium line fluxes (relative to H$\beta$) which are prominently seen in novae spectra, and generate a database.
We explain how these line ratios can be used to estimate the parameters associated with the system, e.g. $T_{BB}$, $L$, and $n_H$ using information derived from observed spectra. We would like to stress here that it is very difficult to get clues about these parameters directly from observations. 
Our paper is organized as following; the modeling procedure is discussed in details in Section \ref{section2}, results are discussed in Section \ref{section3} and summary and conclusions are given in Section \ref{section4}.

\section{Novae Grid Calculation}
\label{section2}
As a first step, we have constructed the grid model using basic assumptions. 
We have considered here dust-free novae as most of the novae have not been observed to form dust. For the dust-forming novae, also, dust form after few weeks, when the temperatures of silicate and graphite grain go below their respective sublimation temperatures.
Here we are primarily interested in novae characteristics in their early phase, where more observations are available. So, we have considered dust-free novae for the calculation. 
Further, to limit computational time and to set a basic abundances, 
we have restricted ourselves to solar metallicity as an average metallicity.
It is possible that the model has limitations, as it is based on simple assumptions, yet we show the method greatly aids in a reliable estimation of the novae parameters.
We are working to construct grid models incorporating dust and higher metallicity which will be presented in a future paper.\\
\\
We use the photoionization code, CLOUDY (version c17.00 rc1) (Ferland et al., 2017), to generate
synthetic spectra of several novae. CLOUDY is based on a self-consistent {\it ab-initio} calculation of the thermal, ionization, and chemical balance. It uses a
minimum number of input parameters and generates output spectra or vice-versa.
Previously, CLOUDY was used in determining elemental abundances and physical
characteristics of few individual nova, such as LMC 1991 (Schwarz et al. 2001), QU Vul (Schwarz  2002), V1974Cyg (Vanlandingham et al. 2005), 
V838 Her $\&$ V4160 Sgr (Schwarz et al.
2007a), V1186 Sco (Schwarz et al. 2007b), V1065 Cen (Helton et al. 2010) and RS Oph (Mondal et al. 2018, Das $\&$
Mondal 2015). Following a similar strategy, we plan to generate a grid of novae models to predict
physical parameters constrained by observed hydrogen and helium line intensity ratios.
For our novae grid models, we consider a spherically expanding ejecta illuminated by
central WD. We assume the central source engine to be a blackbody with surface temperature $T_{BB}$
(in K) and luminosity $L$ (in erg s$^{-1}$). Our calculations include the effects of important ionization
(photo, Auger, collisional, charge transfer) and recombination processes (radiative, dielectronic,
three-body recombination, charge transfer).\\
\\
In addition to this, we assume that half of the
radiation field emitted by the central object actually strikes the gas. Dimensions of spherical
ejecta are defined by inner and outer radii ($R_{in}$, $R_{out}$ ), and the density of the ejecta is set by total
hydrogen density given by
$$ n(H) = n(H^0) + n(H^+) + 2n(H_2) + \sum_{other} n(H_{other}) cm^{-3} ,$$
where $n(H_{other}$) represents H in all other hydrogen-bearing molecules. 
Following Bath \& Shaviv (1976), we assume a radius dependent power-law density profile with exponent $\alpha$, ($ n(R) \propto R^{\alpha}$), $ n(R)$ being the density of the ejecta. Starrfield (1989) had argued that, for novae photospheres, the value of this exponent can only be -2 or -3. A constant mass loss rate and a constant velocity for the ejecta gives rise to a value of -2. However, we assume a constant mass loss rate together with a velocity proportional to the radius from the source. This gives rise to a value of -3 for the exponent, and we use this value in all our models calculated here.
$$n(R) = n(R_{in})\left(\frac{R}{R_{in}}\right)^{\alpha},			\quad \quad				\alpha = -3.$$
Here, $n(R_{in}$) is the density at the illuminated face of the cloud at $R_{in}$. In previous calculations by various authors (e.g. Schwarz et al. 1997, Schwarz et al. 2007b, Helton et al. 2010, Schwarz 2002, etc.) the value of $\alpha$ was also chosen as -3.
We consider a clumpy medium with the filling factor 0.1 and vary inner radius of the ejecta ($R_{in}$),
thickness of ejecta ($\Delta R$), temperature of the central source ($T_{BB}$), luminosity of the central source ($L$),
and hydrogen density ($n_H$). All the models are calculated using solar metallicity.\\
\\
\begin{table}
\begin{center}
\caption{List of the parameters and range of their values considered for the grid model.}
\centering
\begin{tabular}{l c c c c c c c c}
\hline
\hline
\noalign{\smallskip}
Parameters									& Unit					& Range 												& Step size \\
\noalign{\smallskip}
\hline
\noalign{\smallskip}
Inner Radius ($R_{in}$)		     	& cm						& 13.5 $\le$ log($R_{in}$) $\le$ 15.0		& 0.5 \\
Thickness ($\Delta R$)				& cm						& 13.5 $\le$ log($\Delta R$)  $\le$ 15.0		& 0.5 \\
Temperature ($T_{BB}$)			& K					    & 4.5 $\le$ log($T_{BB}$)  $\le$ 6.0		& 0.5 \\
Luminosity	($L$)							& erg s$^{-1}$		& 36.0 $\le$ log($L$)  $\le$ 39.0				& 1.0 \\
H-density ($n_H$)						& cm$^{-3}$			& 6.0 $\le$ log($n_H$)  $\le$ 12.0			& 1.0 \\
\noalign{\smallskip}\hline
\hline
\label{parameter-values}
\end{tabular}
\end{center}
\end{table}
A wide range of values for each of these above mentioned parameters are considered and
the limiting values are chosen on the basis of the available observational results of various classical and recurrent novae in order to construct the grid model.
For example, inner and outer radii are calculated by multiplying 
\onecolumn
\begin{figure}
\centering
\includegraphics[width=6.5 in, height =8 in,clip]{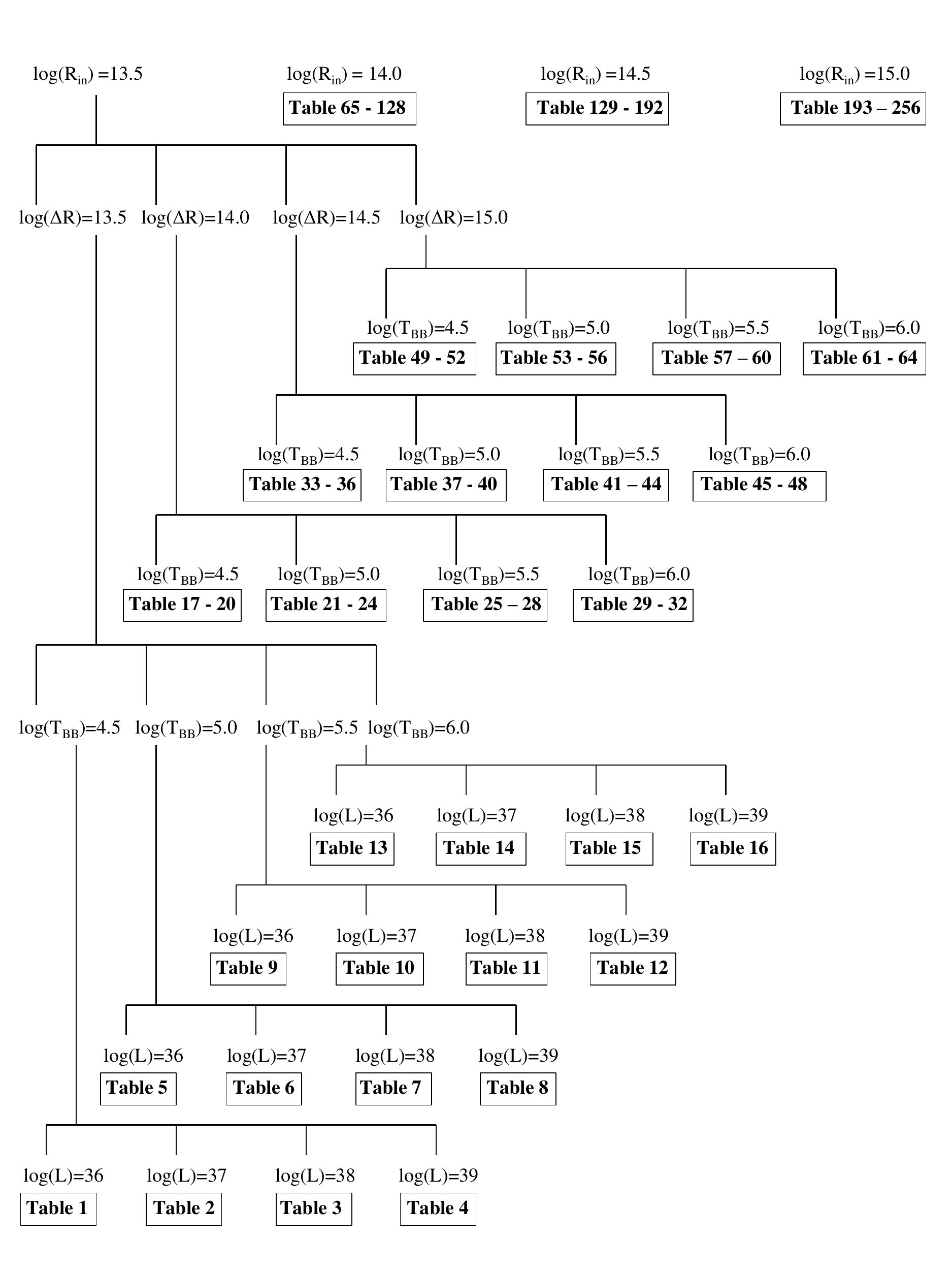}
\caption{Schematic diagram of the structure of datasets of novae grid models. Each table contains data for log($n_H$) = 6 - 12 [in cm$^{-3}$] with step size = 1. See section \ref{section2} \& \ref{section3} for details.}
\label{grid-table}
\end{figure}
\begin{table*}
\begin{center}
\caption{Normalized fluxes of hydrogen and helium emission lines w.r.t. H$\beta$ for log($R_{in})=15.0$ (in cm), log($\Delta R)=14.0$ (in cm), log($T_{BB})=5.5$ (in K), log($L)=36$ (in erg s$^{-1}$), with log($n_H$) = 6 - 12 (in cm$^{-3}$) with step size = 1, corresponding to Table 217 of the database (see Fig. \ref{grid-table}).}
\setlength{\tabcolsep}{2pt}
\clearpage
\centering
\begin{tabular}{>{\centering}p{0.8cm}>{\centering} p{3cm}p{1.8cm}p{1.8cm} p{1.8cm} p{1.8cm} p{1.8cm}p{1.8cm} p{1.8cm}}
\hline
\hline
\multirow{2}{*}{Line ID}&\multirow{2}{*}{Wavelength}&&&&\multirow{2}{*}{log($n_H$) \quad [cm$^{-3}$]}&&&\\
\\
\cline{3-9}
\noalign{\smallskip}
    &$\lambda$ ($\mu$m)&\quad\quad 6&\quad\quad 7&\quad\quad 8&\quad\quad 9&\quad\quad 10&\quad\quad 11&\quad\quad 12\\
\noalign{\smallskip}\hline
\noalign{\smallskip}
Ly C 	&	0.0912	&	2.9190E+01	&	3.8110E+00	&	1.1920E-04	&	0.0000E+00	&	0.0000E+00	&	0.0000E+00	&	1.0110E-12	\\
Ly 6 	&	0.0938	&	4.2240E-01	&	1.0540E-02	&	8.6260E-04	&	8.2940E-04	&	9.0570E-04	&	9.0190E-04	&	9.8810E-04	\\
Ly $\delta$ 	&	0.0950	&	5.8950E-01	&	1.2230E-02	&	1.0530E-03	&	9.0530E-04	&	9.2350E-04	&	8.0760E-04	&	8.3480E-04	\\
Ly $\gamma$ 	&	0.0973	&	7.2740E-01	&	1.2930E-02	&	1.6130E-03	&	1.2480E-03	&	1.1170E-03	&	8.8430E-04	&	7.7610E-04	\\
Ly $\beta$ 	&	0.1026	&	1.0550E+00	&	1.7060E-02	&	3.5310E-03	&	3.0670E-03	&	2.5570E-03	&	1.8120E-03	&	1.3320E-03	\\
Ly $\alpha$	&	0.1216	&	6.7540E+01	&	6.1440E+01	&	1.1850E+02	&	1.2060E+02	&	1.0100E+02	&	7.6470E+01	&	5.3090E+01	\\
H C 	&	0.3646	&	1.6910E+00	&	1.4870E+00	&	1.3630E+00	&	1.3360E+00	&	1.2080E+00	&	9.4060E-01	&	6.7300E-01	\\
H 7 	&	0.3970	&	1.4510E-01	&	1.6930E-01	&	2.0370E-01	&	2.3040E-01	&	2.5220E-01	&	2.3080E-01	&	2.8630E-01	\\
H $\delta$ 	&	0.4102	&	2.4260E-01	&	2.6100E-01	&	3.0420E-01	&	3.3240E-01	&	3.5400E-01	&	3.8010E-01	&	4.0200E-01	\\
H $\gamma$ 	&	0.4340	&	4.5850E-01	&	4.6670E-01	&	4.9190E-01	&	5.1650E-01	&	5.3250E-01	&	5.5070E-01	&	5.7270E-01	\\
He I 	&	0.4388	&	3.2860E-05	&	3.8380E-04	&	6.3160E-03	&	1.2510E-02	&	1.7170E-02	&	1.7270E-02	&	1.3040E-02	\\
He I 	&	0.4471	&	3.1270E-04	&	3.3640E-03	&	5.0270E-02	&	9.4240E-02	&	1.1710E-01	&	1.0480E-01	&	7.2410E-02	\\
He II 	&	0.4686	&	6.8820E-01	&	6.7630E-01	&	3.7720E-01	&	1.7450E-01	&	5.6540E-02	&	1.3110E-02	&	2.0670E-03	\\
H $\beta$ 	&	0.4861	&	1.0000E+00	&	1.0000E+00	&	1.0000E+00	&	1.0000E+00	&	1.0000E+00	&	1.0000E+00	&	1.0000E+00	\\
He I 	&	0.5016	&	1.2640E-05	&	6.6130E-04	&	3.1700E-02	&	6.1880E-02	&	7.8150E-02	&	7.0440E-02	&	4.9750E-02	\\
H $\alpha$	&	0.6563	&	3.0620E+00	&	2.9570E+00	&	4.1380E+00	&	4.9410E+00	&	5.5690E+00	&	5.8580E+00	&	5.8720E+00	\\
He I 	&	0.6678	&	2.3330E-04	&	2.6540E-03	&	4.1060E-02	&	7.8710E-02	&	9.7380E-02	&	8.4280E-02	&	5.5030E-02	\\
He I 	&	0.7065	&	5.3470E-04	&	5.5940E-03	&	6.7980E-02	&	1.2850E-01	&	1.3610E-01	&	9.7780E-02	&	4.9230E-02	\\
Pa C 	&	0.8204	&	3.5220E-01	&	2.9580E-01	&	2.5590E-01	&	2.4380E-01	&	2.1670E-01	&	1.6840E-01	&	1.2060E-01	\\
Pa 9 	&	0.9229	&	2.3490E-02	&	2.6460E-02	&	3.5670E-02	&	4.5440E-02	&	5.4750E-02	&	5.7910E-02	&	5.6870E-02	\\
Pa 8 	&	0.9546	&	3.2700E-02	&	3.6490E-02	&	4.7920E-02	&	5.8920E-02	&	7.0910E-02	&	7.7140E-02	&	7.7890E-02	\\
Pa $\delta$ 	&	1.0049	&	4.8790E-02	&	5.3080E-02	&	6.9660E-02	&	8.2540E-02	&	9.6060E-02	&	1.0620E-01	&	1.0710E-01	\\
He I 	&	1.0830	&	1.5530E-02	&	1.7610E-01	&	1.9340E+00	&	4.5530E+00	&	6.1130E+00	&	5.4110E+00	&	3.3060E+00	\\
Pa $\gamma$ 	&	1.0938	&	8.1120E-02	&	8.3250E-02	&	1.1070E-01	&	1.2920E-01	&	1.4340E-01	&	1.4940E-01	&	1.5040E-01	\\
He II 	&	1.1627	&	3.4190E-02	&	3.3190E-02	&	1.5790E-02	&	7.1870E-03	&	2.4220E-03	&	5.9830E-04	&	1.0090E-04	\\
He I 	&	1.1969	&	1.3500E-05	&	1.5040E-04	&	2.3670E-03	&	4.5540E-03	&	6.0650E-03	&	6.1390E-03	&	5.0290E-03	\\
He I 	&	1.2527	&	1.1200E-05	&	1.2490E-04	&	2.3650E-03	&	4.1690E-03	&	4.1850E-03	&	2.9790E-03	&	1.6330E-03	\\
Pa $\beta$ 	&	1.2818	&	1.3940E-01	&	1.4170E-01	&	1.9280E-01	&	2.2880E-01	&	2.5230E-01	&	2.5150E-01	&	2.4210E-01	\\
Br 20 	&	1.5192	&	1.4960E-03	&	2.1150E-03	&	3.4490E-03	&	4.5010E-03	&	4.9900E-03	&	4.7170E-03	&	3.7980E-03	\\
Br 19 	&	1.5260	&	1.7180E-03	&	2.3770E-03	&	3.9180E-03	&	5.1640E-03	&	5.7750E-03	&	5.4910E-03	&	4.4410E-03	\\
Br 18 	&	1.5342	&	1.9930E-03	&	2.6890E-03	&	4.4630E-03	&	5.9460E-03	&	6.7180E-03	&	6.4340E-03	&	5.2330E-03	\\
Br 17 	&	1.5439	&	2.3350E-03	&	3.0680E-03	&	5.0950E-03	&	6.8670E-03	&	7.8520E-03	&	7.5870E-03	&	6.2150E-03	\\
Br 16 	&	1.5556	&	2.7680E-03	&	3.5450E-03	&	5.8320E-03	&	7.9500E-03	&	9.2170E-03	&	9.0030E-03	&	7.4390E-03	\\
Br 15 	&	1.5701	&	3.3260E-03	&	4.1590E-03	&	6.7000E-03	&	9.2230E-03	&	1.0860E-02	&	1.0740E-02	&	8.9730E-03	\\
Br 14 	&	1.5880	&	4.0580E-03	&	4.9650E-03	&	7.7390E-03	&	1.0730E-02	&	1.2830E-02	&	1.2890E-02	&	1.0890E-02	\\
Br 13 	&	1.6109	&	5.0410E-03	&	6.0400E-03	&	9.0360E-03	&	1.2540E-02	&	1.5190E-02	&	1.5520E-02	&	1.3280E-02	\\
Br 12 	&	1.6407	&	6.3920E-03	&	7.5050E-03	&	1.0760E-02	&	1.4790E-02	&	1.8010E-02	&	1.8700E-02	&	1.6190E-02	\\
Br 11 	&	1.6806	&	8.3030E-03	&	9.5620E-03	&	1.3180E-02	&	1.7700E-02	&	2.1390E-02	&	2.2520E-02	&	1.9680E-02	\\
He I 	&	1.7003	&	2.2110E-05	&	2.3790E-04	&	3.5560E-03	&	6.6670E-03	&	8.2830E-03	&	7.4150E-03	&	5.3760E-03	\\
Br 10 	&	1.7362	&	7.6950E-03	&	8.7790E-03	&	1.2330E-02	&	1.5620E-02	&	1.8300E-02	&	1.8710E-02	&	1.8320E-02	\\
Br 9 	&	1.8174	&	1.0450E-02	&	1.1330E-02	&	1.5200E-02	&	1.9340E-02	&	2.3300E-02	&	2.4640E-02	&	2.4920E-02	\\
Pa $\alpha$ 	&	1.8751	&	2.8440E-01	&	2.7860E-01	&	4.1780E-01	&	5.1640E-01	&	5.7850E-01	&	5.7250E-01	&	5.3670E-01	\\
Br $\delta$ 	&	1.9446	&	1.4910E-02	&	1.5680E-02	&	2.0380E-02	&	2.4990E-02	&	3.0050E-02	&	3.2680E-02	&	3.3900E-02	\\
He I 	&	2.0581	&	7.0750E-07	&	1.7650E-04	&	4.1300E-02	&	1.2850E-01	&	2.3440E-01	&	2.6670E-01	&	1.9320E-01	\\
He I 	&	2.1130	&	2.7780E-06	&	2.9060E-05	&	3.8350E-04	&	6.8760E-04	&	7.6670E-04	&	5.8670E-04	&	3.4470E-04	\\
Br $\gamma$ 	&	2.1655	&	2.2750E-02	&	2.3090E-02	&	2.9590E-02	&	3.4830E-02	&	4.0430E-02	&	4.4670E-02	&	4.6070E-02	\\
Br $\beta$ 	&	2.6252	&	3.8300E-02	&	3.6890E-02	&	4.6910E-02	&	5.3890E-02	&	5.9490E-02	&	6.1790E-02	&	6.3500E-02	\\
Pf 10 	&	3.0384	&	4.0130E-03	&	4.5100E-03	&	6.3250E-03	&	8.0140E-03	&	9.3860E-03	&	9.5950E-03	&	9.3460E-03	\\
Pf 9 	&	3.2961	&	5.5080E-03	&	5.8080E-03	&	7.7740E-03	&	9.8810E-03	&	1.1900E-02	&	1.2580E-02	&	1.2640E-02	\\
Pf $\gamma$ 	&	3.7400	&	7.9990E-03	&	8.0270E-03	&	1.0360E-02	&	1.2680E-02	&	1.5240E-02	&	1.6560E-02	&	1.7150E-02	\\
Br $\alpha$ 	&	4.0512	&	6.3160E-02	&	5.9880E-02	&	7.7720E-02	&	9.0410E-02	&	9.8890E-02	&	9.8270E-02	&	9.6580E-02	\\
Pf $\beta$ 	&	4.6525	&	1.2430E-02	&	1.1780E-02	&	1.4840E-02	&	1.7390E-02	&	2.0160E-02	&	2.2250E-02	&	2.3070E-02	\\
Hu $\gamma$ 	&	5.9066	&	3.2200E-03	&	3.3250E-03	&	4.4410E-03	&	5.6420E-03	&	6.7920E-03	&	7.1820E-03	&	7.2460E-03	\\
Pf $\alpha$ 	&	7.4578	&	2.1030E-02	&	1.8280E-02	&	2.2390E-02	&	2.5420E-02	&	2.7950E-02	&	2.8990E-02	&	3.0160E-02	\\
Hu $\beta$ 	&	7.5004	&	4.7050E-03	&	4.5330E-03	&	5.8230E-03	&	7.1170E-03	&	8.5480E-03	&	9.2910E-03	&	9.6840E-03	\\
Hu $\alpha$ 	&	12.3685	&	7.2340E-03	&	6.3200E-03	&	7.8650E-03	&	9.1900E-03	&	1.0640E-02	&	1.1740E-02	&	1.2300E-02	\\
\noalign{\smallskip}\hline
\hline
\label{grid-sample-flux}
\end{tabular}
\clearpage
\end{center}
\end{table*}
\twocolumn
velocities (which can be calculated from the line-widths) with time and the thickness can be calculated by subtracting the inner radius from outer one. 
The upper and lower limits of radii are determined considering high velocity of novae ejecta of 10,000 km/s (e.g. U Sco, Banerjee et al. 2010)
and lower expanding value of 300 km/s (e.g. V723 Cas, Iijima, 2006). Since emission lines generally start to get resolved about a week after outburst, in the present calculation, we consider time from day 5 and calculate upto day 120 after outburst. On the basis of above values, the limiting values of log($R_{in}$) (in cm) and log( $\Delta R$) (in cm) are chosen as 13.5 and 15.0 with step-size 0.5. Similarly, we determine the limiting values of other parameters, namely, temperature, luminosity, and hydrogen density on the basis of available observational results. For temperature($T_{BB}$) (in K) and luminosity($L$) (in erg s$^{-1}$) of the ionizing source, we have considered log($T_{BB}$) $\in$ [4.5, 6.0] with step-size 0.5 and log($L$) $\in$ [36.0, 39.0] with step-size 1.0, respectively. Finally, we have taken log($n_H$) (in cm$^{-3}$) $\in$ [6.0, 12.0] with step-size 1.0. Details of the parameter values are presented in tabular form in Table \ref{parameter-values}. Our calculations stop at the thickness ($\Delta R$) taken for a particular model. Thus a total of 1792 models were constructed by varying each of these parameters. 
The results i.e. ratio of line fluxes are arranged in
256 tables. These 256 tables of dataset are available at the website {\it https://aninditamondal1001.wixsite.com/researchworks} under
the button ``\textsc{Novae Grid Data}" on the page \textsc{\it Research/List-of-Documents}. Structure of the tables are briefly shown in Fig. \ref{grid-table}.

\section{Results}
\label{section3}
\begin{figure}
\centering
\includegraphics[width=3.4 in, height =3.4 in, clip]{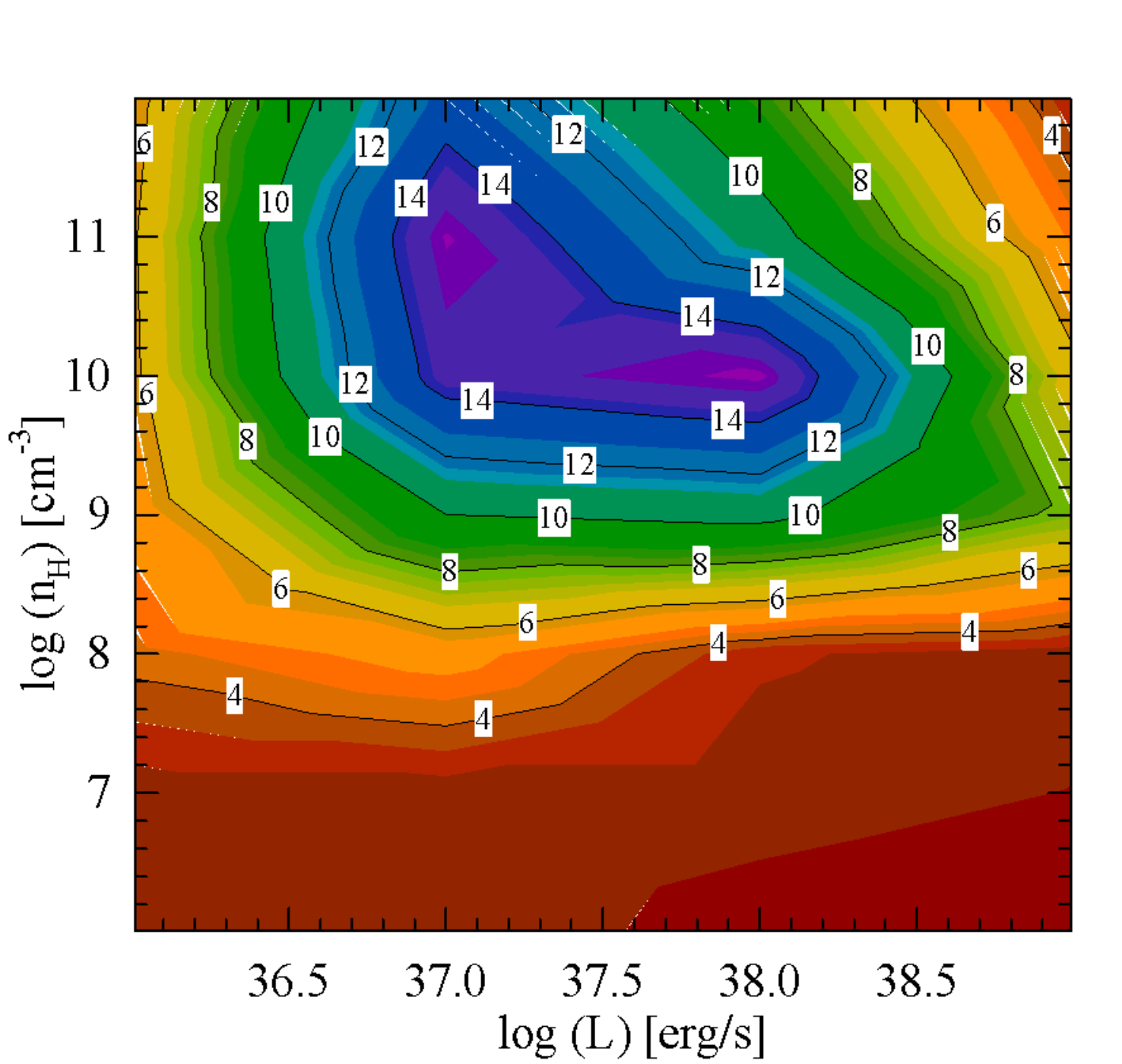}
\caption{Contour plot of H$\alpha$ line flux ratio w.r.t. H$\beta$ for log($R_{in}$) = 15.0 (in cm), log({$\Delta R$}) = 14.0 (in cm), and
log($T_{BB}$) = 5.5 (in K); log($L$) (in erg s$^{-1}$) and log($n_H$) (in cm$^{-3}$) are plotted along x and y-axes respectively, corresponding to the Tables 217 - 220 of database (see Fig. \ref{grid-table}). The values of H$\alpha$/H$\beta$ are mentioned on each contour. See section \ref{section3} for more details.}
\label{contour}
\end{figure}

\begin{figure}
\centering
\includegraphics[width=3.4 in, height= 3.4 in, clip]{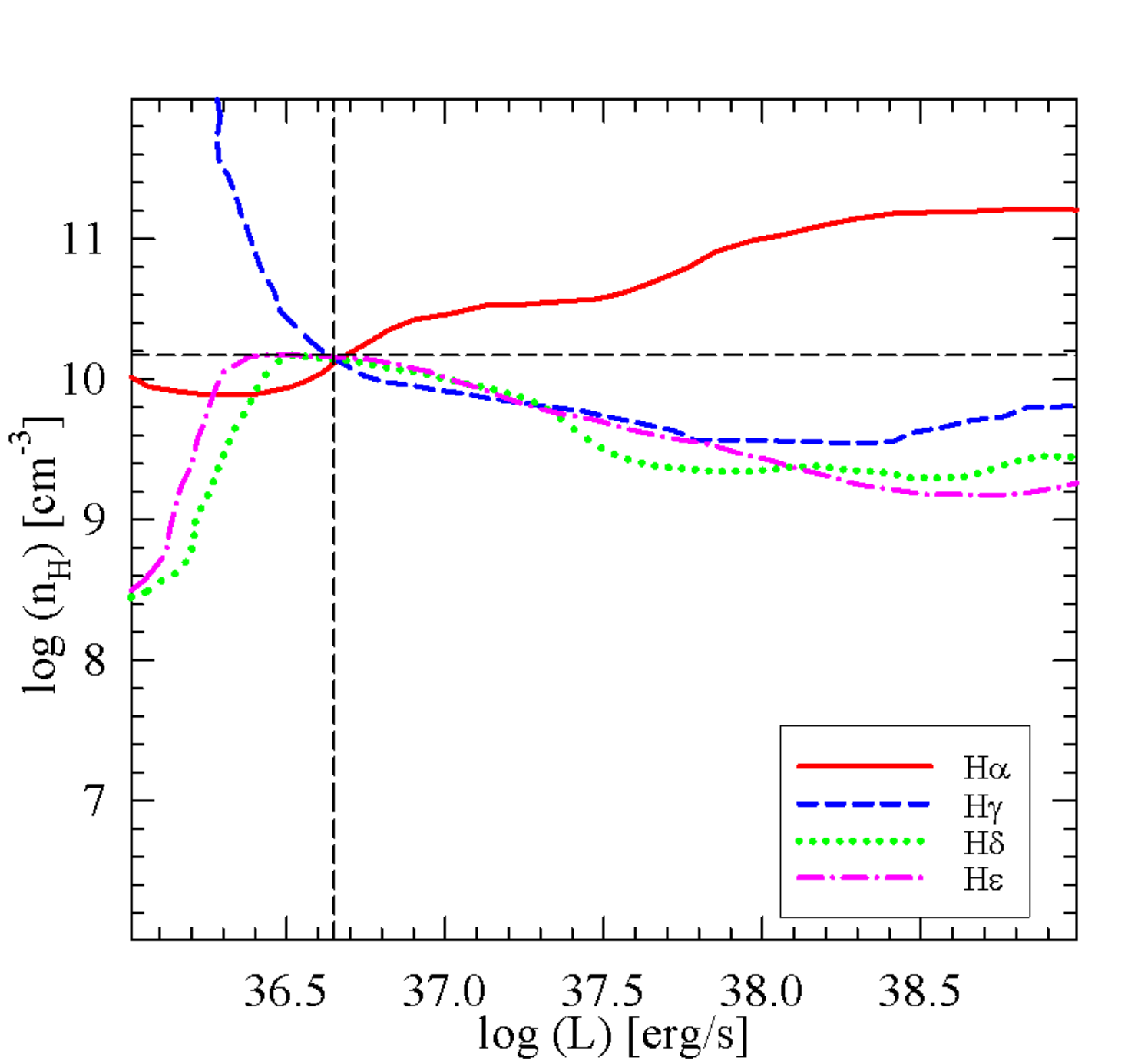}
\caption{Plots of different extracted contour lines for nova RS Oph, 12 days after outburst. The
red solid, blue short dashed, green dotted, and pink dotted dashed
lines represent the extracted plots of line flux ratios of H$\alpha$, H$\gamma$, H$\delta$ $\&$ H$\epsilon$ respectively w.r.t. H$\beta$. The lines intersect at log($L$) = 36.65 (in erg s$^{-1}$) and log($n_H$) = 10.2 (in cm$^{-3}$). See section \ref{section3} for more details.}
\label{RSOph-grid}
\end{figure}

From the model generated spectra, we have calculated the fluxes of fifty-six hydrogen and helium recombination lines which are generally seen prominently in novae emission spectra, covering a wide wavelength region.
As an example, we present the results of such a model for log($R_{in}$) = 15.0 (in cm), log($\Delta R$) = 14.0 (in cm), log($T_{BB}$) = 5.5 (in K) and log($L$) = 36 (in erg s$^{-1}$) in Table \ref{grid-sample-flux} (corresponding to Table 217 of database). Line fluxes have been calculated for 7 different values of log($n_H$) (in cm$^{-3}$) (i.e. for $n_H$ = $10^6, 10^7, 10^8, 10^9, 10^{10}, 10^{11}$ and $10^{12}$). 
Contours of line flux ratios (along z-axis) can be plotted against any two parameters (among $T_{BB}$, $L$ and $n_H$) along x and y-axes, keeping the other one fixed. 
A typical contour plot of H$\alpha$ line flux ratio is shown in Fig. \ref{contour}, with log($L$) along the x-axis and log($n_H$) along the y-axis, for log($R_{in}$) = 15.0  (in cm), log($\Delta R$) = 14.0 (in cm), and log($T_{BB}$) = 5.5 (in K), corresponding to the Tables 217 - 220 of the database (see Fig. \ref{grid-table} and Table \ref{grid-sample-flux}). The values of line flux ratios of H$\alpha$ w.r.t. H$\beta$ are mentioned on the contours.\\
\\
As mentioned earlier in section \ref{section2}, $R_{in}$ and {$\Delta R$} could be calculated from line-widths and time elapsed after outburst; $T_{BB}$ could be obtained by blackbody fitting with the continuum of the observed spectra and line fluxes could be measured from the observed spectra. Once, the values of $R_{in}$, $\Delta R$, and $T_{BB}$ are found, we can choose the corresponding dataset from the database and make contour plots for different lines, e.g. H$\alpha$, H$\gamma$, H$\delta$ etc.  Now, from these contour plots we can extract (e.g. Fig. \ref{contour}) the contours for the corresponding values of the line ratios and plot them together (e.g. Fig. \ref{RSOph-grid}). From the intersection of the contours of different lines, we can determine the values of the parameters viz. $L$ and $n_H$ (in case of Fig. \ref{RSOph-grid}). In this way, by knowing the value of any one of $L$, $n_H$, and $T_{BB}$, values of the other two could be determined.\\ 
\\
\begin{table*}
\begin{center}
\caption{Comparison of estimated parameter values of few novae obtained from grid model with previously calculated results (shown in paranthesis)}
\centering
\begin{tabular}{l c c cp{1.5 cm} p{1.5 cm} cp{1.5 cm} cp{1.5 cm} cp{1.5 cm}}
\hline
\hline
\noalign{\smallskip}
\multirow{2}{*}{Novae}&\multirow{2}{*}{Line flux ratio}&\multirow{2}{*}{Ref.}&&\multirow{2}{*}{Estimated values (previously calculated values)}&&&\\
\\
\cline{4-8}
\noalign{\smallskip}
    (Outburst)&  from literature &  & log($R_{in}$) & log($\Delta R$) & log($T_{BB}$)  & log($L$) & log($n_H$)\\
    &&& [in cm]&[in cm]& [in K]& [in erg s$^{-1}$]& [in cm$^{-3}$]\\
\cline{1-8}
\noalign{\smallskip}
RS Oph				 	& H$\alpha$/H$\beta$ = 6.12, H$\gamma$/H$\beta$ = 0.39,  	&  1& 14.0 (14.0$^1$)	& 14.0 (14.05$^1$) 	& 4.5 (4.5$^1$)	& 36.65 (36.8$^1$) & 10.2 (10.5$^1$) \\
(2006)																& H$\delta$/H$\beta$ = 0.28, H$\epsilon$/H$\beta$ = 0.29 &&&&& \\
\noalign{\smallskip}
V1065 Cen				& H$\alpha$/H$\beta$ = 3.89, H$\gamma$/H$\beta$ = 0.48,  &  2 & 15.0 (15.16$^2$)	& 15.0 (15.02$^2$) 	& 5.0 (4.77$^2$)& 38.0 (38.05$^2$)& 7.5 (7.52$^2$)	 \\
(2007)																& Pa$\gamma$/Pa$\beta$ = 0.61  &&&&&\\
\noalign{\smallskip}
V1186 Sco				&	 H$\gamma$/H$\beta$ = 0.38, H$\delta$/H$\beta$ = 0.24,  & 3& 15.0 (15.04$^3$)	& 15.0 (15.15$^3$) 	& 4.5 (4.7$^3$) & 36.8 (36.8$^3$)& 7.5 (7.5$^3$)  \\
(2004)																& Pa$\gamma$/Pa$\beta$ = 0.57, Hu$\gamma$/Hu$\alpha$ = 0.65, &&&&&\\
																		& He I (1.083 $\mu$m)/Pa$\beta$ = 12.14  &&&&&\\
\noalign{\smallskip}
V1974 Cyg 			& H$\alpha$/H$\beta$ = 2.93, H$\gamma$/H$\beta$ = 0.42 	&  4 & 15.0 (15.3$^4$)& 15.0 (15.4$^4$) 	& 5.5 (5.52$^4$) & 38.0 (38.06$^4$)& 8.0 (7.8$^4$) \\
(1992)&&&&\\
\noalign{\smallskip}
PW Vul				& H$\alpha$/H$\beta$ = 4.4, H$\gamma$/H$\beta$ = 0.37, 	& 5& 15.0 (15.25$^5$)	& 15.0 (15.5$^5$) 	& 5.5 (5.4$^5$) & 37.7 (37.8$^5$)& 7.1 (7.04$^5$) \\
(1984)																& He II (4686 $\AA$)/H$\beta$ = 0.34  &&&&&\\
\noalign{\smallskip}\cline{1-8}
\cline{1-8}
\label{other-cloudy-results}
\end{tabular}
\begin{tablenotes}
$^1$Mondal et al. 2018, Das $\&$ Mondal 2015;  $^2$Helton et al. 2010;  $^3$Schwarz et al. 2007b;  $^4$Vanlandingham et al. 2005;  $^5$Schwarz et al. 1997.
\end{tablenotes}
\end{center}
\end{table*}
\begin{figure}
\centering
\includegraphics[width=3.65 in, height= 3.1 in, clip]{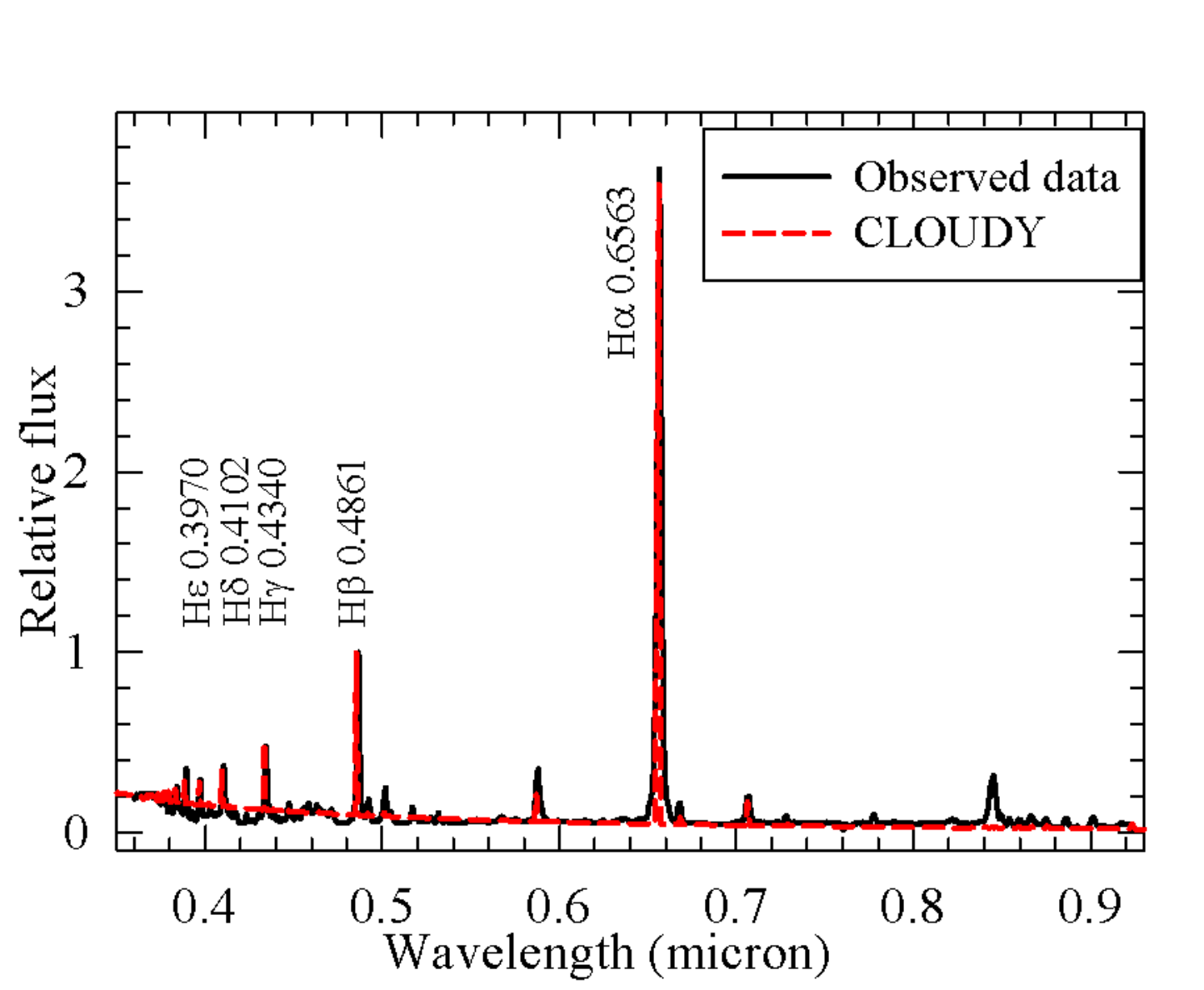}
\caption{Comparison of observed (solid black) and CLOUDY generated (dashed red) spectra of nova RS Oph. Prominent hydrogen features are marked. See section \ref{section3} for more details.}
\label{RSOph-spectra-grid}
\end{figure}
To check if this method works well,
we apply this method to few novae. First, we consider the example of RS Ophiuchi (RS Oph), which is a well-known recurrent nova (recurrence period $\sim$ 20 years). 
From the optical spectra taken 12 days after outburst (2006) with 2m Himalayan Chandra Telescope (HCT), we have calculated the temperature as 10$^{4.5}$ K
and the line flux ratio of H$\alpha$, H$\gamma$, H$\delta$ $\&$ H$\epsilon$ w.r.t. H$\beta$ as 6.12, 0.39, 0.28 $\&$ 0.29 respectively (Mondal et al. 2018).
From the expansion velocities of the ejecta (see section 5 in Mondal et al., 2018) we have calculated $R_{in}$ and $\Delta R$. For this particular set of $R_{in}$, $\Delta R$, and $T_{BB}$ we have plotted contours for the hydrogen lines with $L$ and $n_H$ along x and y-axes. We have extracted the contours for the corresponding values of the line ratios and have plotted them together (Fig. \ref{RSOph-grid}). From the figure, we find that all lines intersect at log($L$) = 36.65 (in erg s$^{-1}$) and log($n_H$) = 10.2 (in cm$^{-3}$). These values are consistent with previous results of RS Oph (see Table \ref{other-cloudy-results}).\\
\\
We also run a seperate CLOUDY model for RS Oph considering the derived parameter values and have generated synthetic spectrum keeping the abundances as solar. The simulated spectra was then compared with the observed one taken in 2m HCT, in the optical region. In Fig. \ref{RSOph-spectra-grid}, the observed spectra is shown in solid black line and the simulated spectra is shown in dashed red line. It is clearly seen that the observed hydrogen features are matching well with the modelled hydrogen lines. This, alternatively, validate our method of estimating parameters.\\
\\
In a similar way, we have also tested our grid model for other galactic classical novae e.g. V1065 Cen, V1186 Sco, V1974 Cyg and PW Vul to determine the physical parameters. For novae RS Oph, V1065 Cen, and V1974 Cyg, we could estimate the parameter values using only hydrogen lines. But, in case of novae V1186 Sco and PW Vul, contours of hydrogen lines (optical, NIR and FIR) intersected at multiple points. So, we use the He I 1.083 $\mu$m line (for V1186 Sco) and He II 4686 $\AA$ line (for PW Vul) which are strong in the spectra, along with other hydrogen lines to estimate the parameters.
It may be mentioned here that from previous studies, we find the He abundances w.r.t. solar as 1.1 $\pm$ 0.3 in case of V1186 Sco (Schwarz et al., 2007b) and  1.0 $\pm$ 0.4 in case of PW Vul (Schwarz et al., 1997), which are near to solar values. 
So, there are no significant changes in the intensities of He lines and use of these lines are quite safe to estimate the parameters.
Our results match well with the previously calculated results; the results are shown in Table \ref{other-cloudy-results}.
These results validate the method and motivate us to apply this to few other novae and estimate the physical parameter values.\\
\\
\begin{figure}
\centering
\includegraphics[width=3.4 in, height =3.4 in, clip]{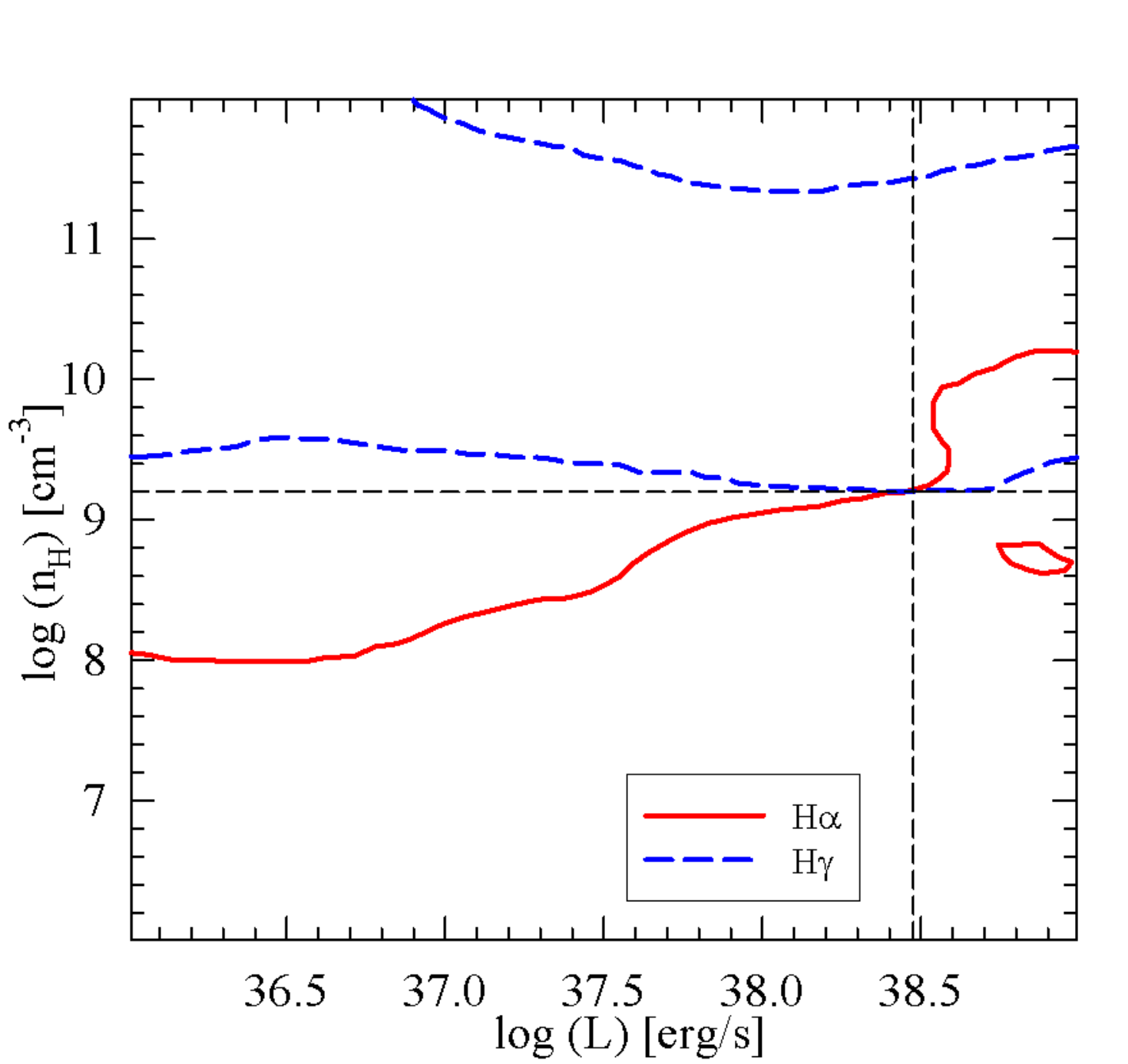}
\caption{Plots of different extracted contour lines of nova KT Eri, 12 days after outburst. The 
red solid and blue dashed lines represent the extraccted plots of line flux ratios of H$\alpha$ and H$\gamma$ respectively w.r.t. H$\beta$. The lines
intersect at log($L$) = 38.5 (in erg s$^{-1}$) and log($n_H$) = 9.2 (in cm$^{-3}$). See section \ref{section3} for more details.}
\label{KTEri-grid}
\end{figure}
\\
(i) \textbf{Nova KT Eridin (KT Eri)}: this is a well known galactic classical nova, discovered by K. Itagaki on 2009 November 25.5 UT (Yamaoka et al. 2009).
The maximum and minimum velocities of the ejecta are determined as $V_{max}$ = 3600 km/s, $V_{min}$ = 1900 km/s respectively, from the broad Balmer emission features (Maehara, Arai \& Isogai 2009). From the SMARTS spectroscopic data (Walter et al., 2012), taken 12 days after outburst, we have calculated $T_{BB}$ = $10^5$ K. The line flux ratio of H$\alpha$ and H$\gamma$ w.r.t. H$\beta$ are measured as 2.87 and 0.78 respectively.
Using the values in the database, we have plotted the contours for the  hydrogen lines for the particular set of $R_{in}$, $\Delta R$, and $T_{BB}$. Then we have extracted contours for the observed line fluxes together (Fig. \ref{KTEri-grid}) and from the intersection of the lines,
we have estimated log($L$) = 38.5 (in erg s$^{-1}$) $\&$ log($n_H$) = 9.2 (in cm$^{-3}$).\\
\\
\begin{figure}
\centering
\includegraphics[width=3.4 in, height =3.4 in, clip]{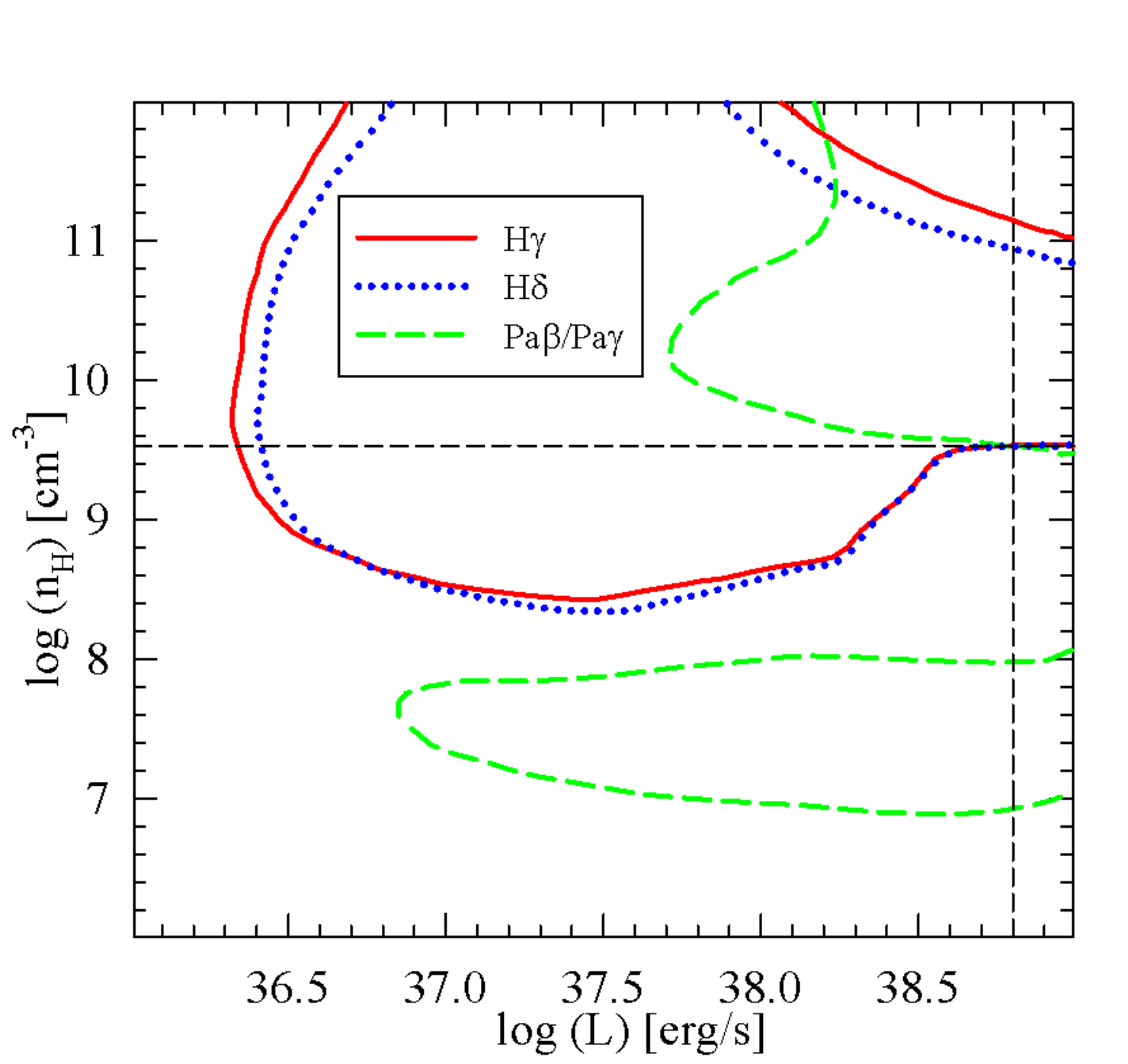}
\caption{Plots of different extracted contour lines of nova V5558 Sgr, 24 days after outburst. The
red solid and blue dotted lines represent the extracted plots of line flux ratios of H$\gamma$ and H$\delta$ respectively w.r.t. H$\beta$ and the
green dashed line represents the extracted plots of line flux ratios of Pa$\beta$ w.r.t. Pa$\gamma$. The lines intersect at log($L$) = 38.8 (in erg s$^{-1}$) and log($n_H$) = 9.5 (in cm$^{-3}$). See section \ref{section3} for more details.}
\label{V5558Sgr-grid}
\end{figure}
\\
(ii) \textbf{V5558 Sagittarii (V5558 Sgr)}: this nova was discovered on 2007 April 14.77 UT. The $R_{in}$ and $\Delta R$ on 24 days after outburst are calculated using the minimum ($V_{min}$ = 250 km/s) and maximum ($V_{max}$ = 540 km/s) expansion velocities of the ejecta (Iijima 2007a, 2007b). From the continuum of the optical spectra, we have measured the temperature to be $10^6$ K. We have calculated the line flux ratio of H$\gamma$ and H$\delta$ w.r.t. H$\beta$ and Pa$\beta$ w.r.t. Pa$\gamma$ from the spectra taken in the optical (Tanaka et al. 2011) and near-infrared region (Das et al. 2015) respectively. We have then plotted the corresponding contours from the database. Then we have extracted the contours for the observed line ratios and have plotted them together (Fig. \ref{V5558Sgr-grid}). From the intersection of the extracted contours for the observed line ratios, we estimate $L =  6.4 \times 10^{38}$  erg s$^{-1}$ and $n_H = 3.4 \times 10^9$ cm$^{-3}$.
This example also shows that lines at other wavelength region may help sometimes to determine the values more precisely. A more detailed study of this nova using CLOUDY is under progress and will appear in a seperate paper in future.\\
\\
\begin{figure}
\centering
\includegraphics[width=3.4 in, height =3.4 in, clip]{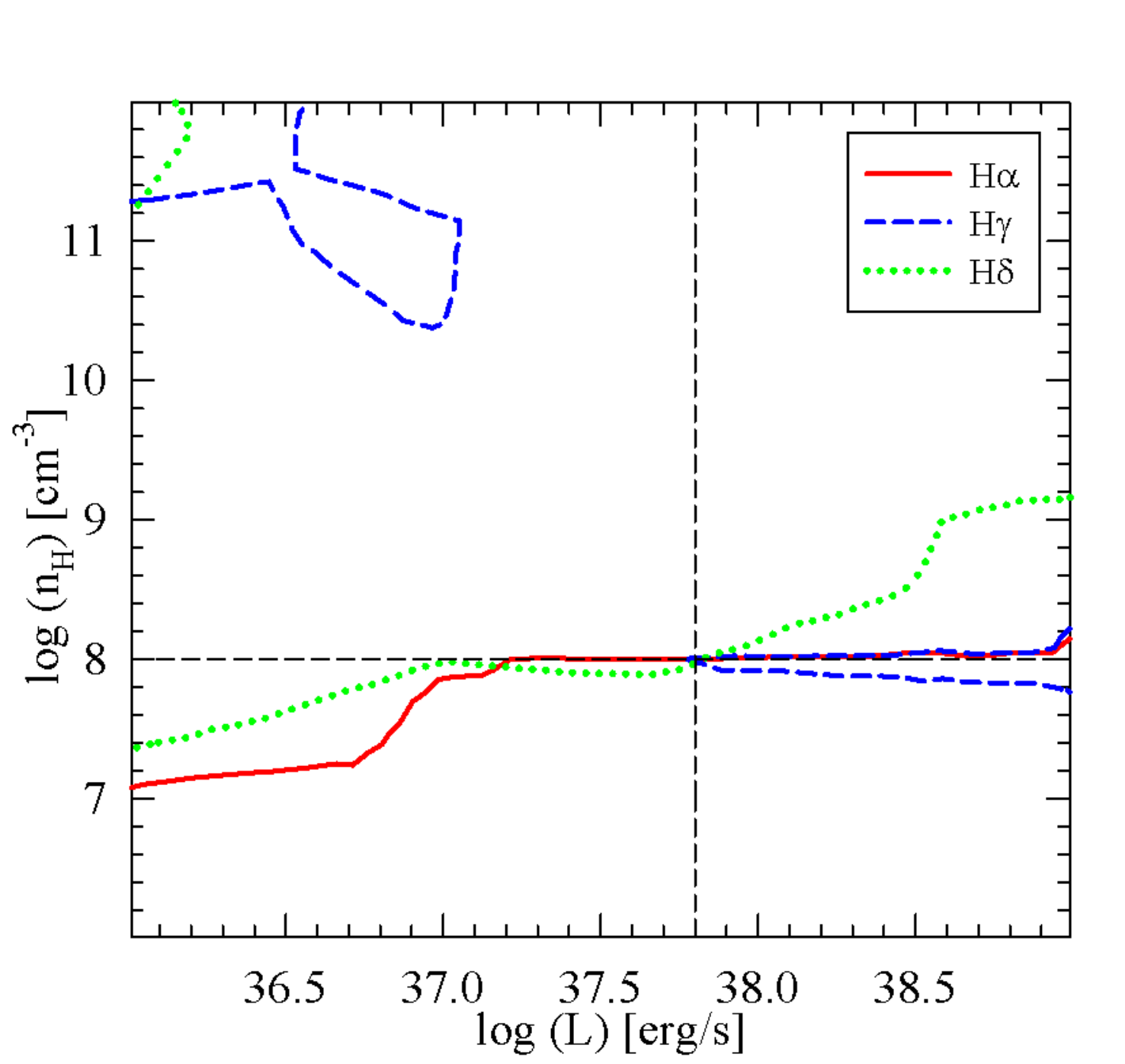}
\caption{Plots of different extracted contour lines of nova U Sco, 5.81 days after outburst. The
red solid, blue short dashed, and green dotted
lines represent the extracted plots of line flux ratios of H$\alpha$, H$\gamma$, $\&$ H$\delta$ respectively w.r.t. H$\beta$. The lines intersect at log($L$) = 37.8 (in erg s$^{-1}$) and log($n_H$) = 8.0 (in cm$^{-3}$). See section \ref{section3} for more details.}
\label{USco-grid}
\end{figure}
\\
(iii) \textbf{U Scorpii (U Sco)}: the recurrent nova U Sco explodes at intervals of 10 $\pm$ 2 years; it's recent outburst occured on Jan 28, 2010 (Schaefer et al. 2010). From the optical spectra taken on about 6 days after outburst, maximun velocity was calculated from the He I 7065 $\AA$ line as $\sim$ 9900 km/s whereas minimun velocity was calculated as $\sim$ 3786 km/s from the hydrogen lines  (Anupama et al., 2013). During the early phase, effective tempertaure was measured as $2 \times 10^4$ K and electron density as $10^{7.8}$ cm$^{-3}$ (Anupama et al. 2013). Line fluxes of H$\alpha$, H$\gamma$, and H$\delta$ w.r.t. H$\beta$ are measured as 2.45, 0.72 and 0.38. Then following the similar method, from extracted contour plots (Fig. \ref{USco-grid}), we have estimated $n_H$ = $10^8$ $cm^{-3}$ and $L$ = $10^{37.8}$ erg s$^{-1}$, which match well with the previously obtained results (see Table \ref{Cloudy-grid-new-resuts}).\\
\\
\begin{figure}
\centering
\includegraphics[width=3.4 in, height =3.4 in, clip]{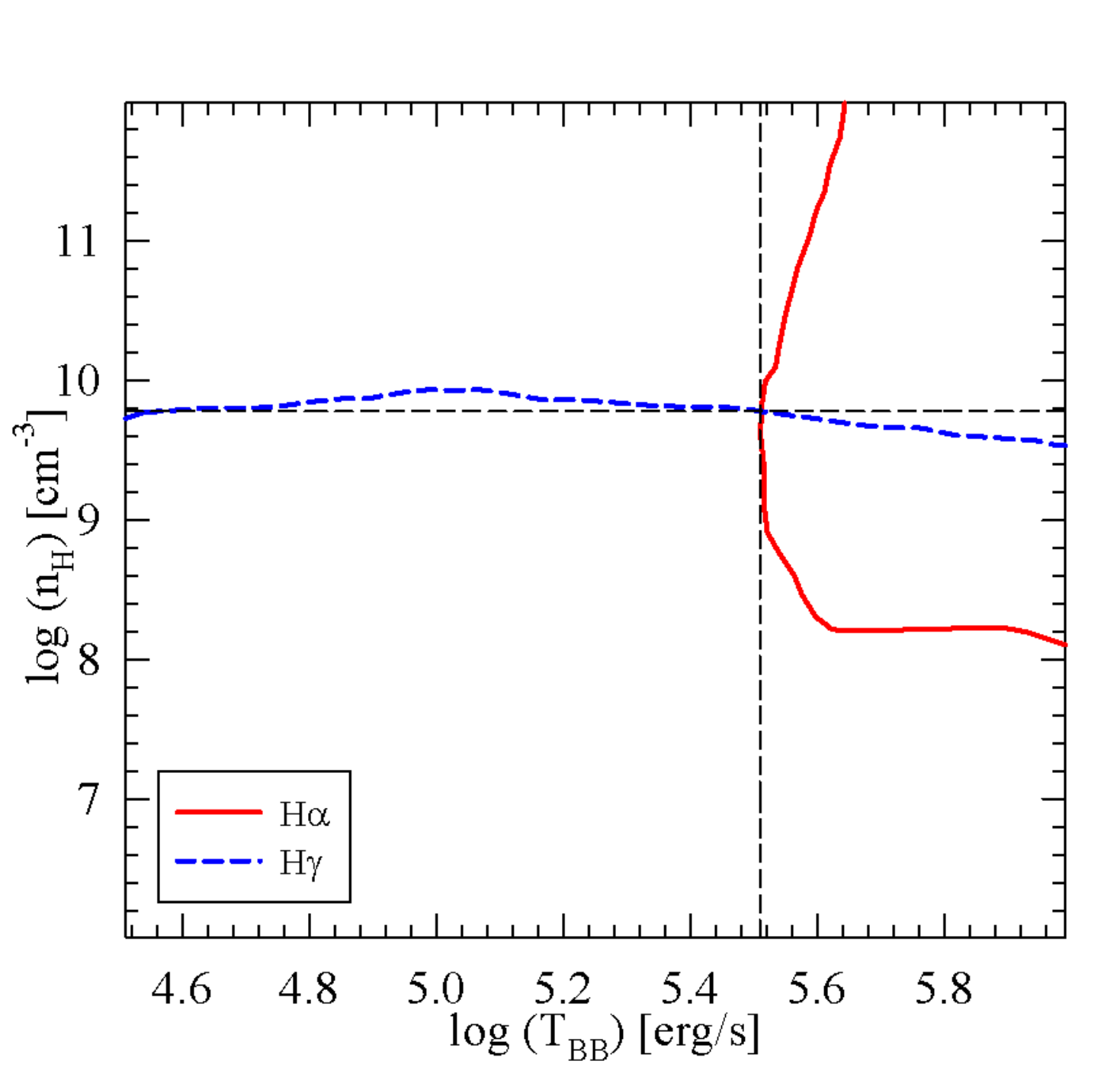}
\caption{Contour plot of extracted contour lines of nova V339 Del, 21 days after outburst. The 
red solid and blue dashed lines represent the contour plots of line flux ratio of H$\alpha$ and H$\gamma$ respectively w.r.t. H$\beta$. The lines intersect at log($T_{BB}$) = 5.51 (in K) and log($n_H$) = 9.78 (in cm$^{-3}$). See section \ref{section3} for more details.}
\label{V3339Del-grid}
\end{figure}
\\
(iv) \textbf{V339 Delphini (V339 Del)}: On 14.584 August, 2013, Koichi Itagaki discovered classical nova Delphini 2013 (V339 Del) at 6.8 optical magnitude (Nakamo et al. 2013). Burlak et al. (2015) calculated the luminosity of the system as $10^{38.39}$ erg s$^{-1}$ from the early phase spectrum and the expansion velocity was calculated in the range of 1000 - 1800 km/s from the H$\alpha$ emission lines. We have measured the Balmer line flux ratios from the spectra taken on 21 days after the outburst as H$\alpha$/H$\beta$ = 6.34 and H$\gamma$/H$\beta$ = 0.34. 
Here, we have drawn the contour plots of hydrogen line flux ratios with $T_{BB}$ and $n_H$ along x and y-axes respectively (Fig. \ref{V3339Del-grid}). The lines intersect at log($T_{BB}$) = 5.51 (in K) and log($n_H$) = 9.78 (in cm$^{-3}$), which give the values of $T_{BB}$ and $n_H$ respectively.\\
\\
\begin{figure}
\centering
\includegraphics[width=3.4 in, height =3.4 in, clip]{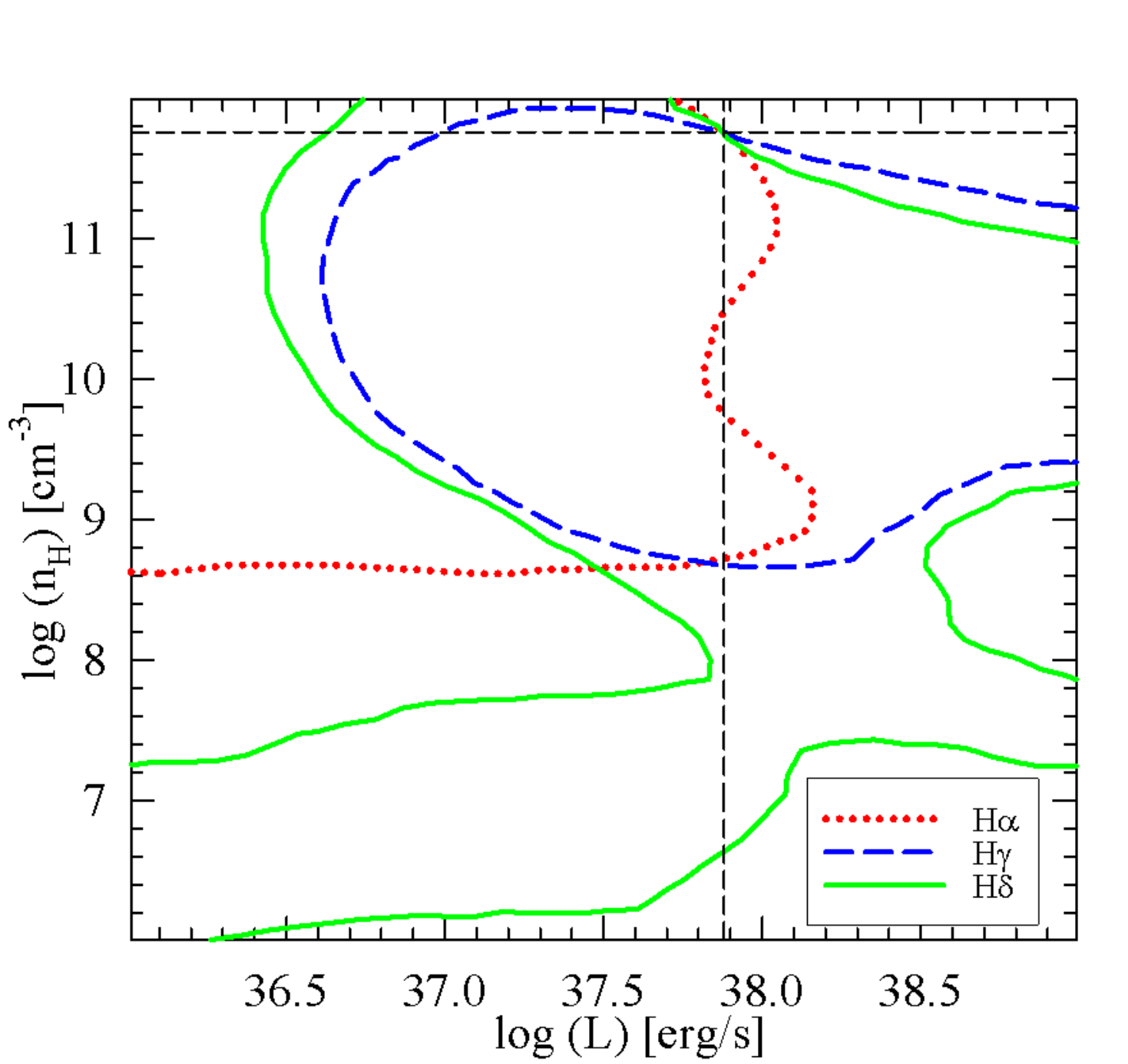}
\caption{Plots of extracted contour lines of nova 2015 of dwarf galaxy IC1613. 26 days after ouburst. The
red dotted, blue dashed, and green solid
lines represent the extracted plots of line flux ratios of H$\alpha$, H$\gamma$, $\&$ H$\delta$ respectively w.r.t. H$\beta$. The results are log($L$) =  37.9 (in erg s$^{-1}$) and log($n_H$) = 8.7 (in cm$^{-3}$). See section \ref{section3} for more details.}
\label{IC1613-grid}
\end{figure}
\\
(v) \textbf{IC 1613 \#2015}: this is an extra-galactic classical nova in dwarf galaxy IC 1613 discovered on 2015 September 10. The V$_{max}$ and V$_{min}$ were calculated from H$\alpha$ emission line and H$\gamma$ absorption line as $\sim$ 1750 km/s and 1200 km/s respectively (Williams et al. in 2017). From the {\it Swift} X-rays studies, temperature and luminosity of the source were determined to be $T_{BB} = 10^{5.76}$ K and $L$ = $10^{37.7}$ erg s$^{-1}$ (Williams  et al. 2017). We have used the line flux ratio of H$\alpha$, H$\gamma$ and H$\delta$ w.r.t. H$\beta$ measured 26 days after outburst to find the values of other observables. The extracted contours of the line flux ratios from the grid database gives log($L$) =  37.9 (in erg s$^{-1}$), which matches well with $L$ = $10^{37.7}$ erg s$^{-1}$ from the X-ray study, and log($n_H$) = 11.7 (in cn$^{-3}$). The plot is shown in Fig. \ref{IC1613-grid} and the parameter values are shown in Table \ref{Cloudy-grid-new-resuts} in more detail.\\
\\

\begin{table*}
\begin{center}
\caption{Estimated parameter values of few novae obtained through CLOUDY grid model}
\centering
\begin{tabular}{l c c c c c c c c}
\hline
\hline\noalign{\smallskip}
Novae &  Outburst  & Line Flux Ratios &  References & log($R_{in}$) & log($\Delta R$) & log($T_{BB}$)  & log($L$) & log($n_H$) \\
&&&&[in cm]&[in cm]& [in K]& [in erg s$^{-1}$]& [in cm$^{-3}$]\\
\hline
\noalign{\smallskip}
KT Eri 					& 2009 	& H$\alpha$/H$\beta$ = 2.87, H$\gamma$/H$\beta$ = 0.78  & 1 & 14.0	& 14.0 	& 5.0 & 38.5 	& 9.2	   \\
\noalign{\smallskip}
V5558 Sgr 				& 2007	& H$\gamma$/H$\beta$ = 0.52, H$\delta$/H$\beta$ = 0.20,  & 2 & 14.5 	& 14.5 	& 6.0 & 38.8 	& 9.53	  \\
																				&& Pa$\beta$/Pa$\gamma$ = 1.14 &&&&&\\
\noalign{\smallskip}
U Sco & 2010	&	H$\alpha$/H$\beta$ = 2.45, H$\gamma$/H$\beta$ = 0.72, & 3 &	14.5	&	14.5	&	4.5	&	37.8	& 8.0	  \\
																				&& H$\delta$/H$\beta$ = 0.38 &&&&&\\
\noalign{\smallskip}																	
V339 Del & 2013 & H$\alpha$/H$\beta$ = 6.34, H$\gamma$/H$\beta$ = 0.34 & 4  & 14.5	&	14.0 &	5.51	& 38.0	& 	9.78	 \\
\noalign{\smallskip}
Nova IC1613 2015	& 2015	& H$\alpha$/H$\beta$ = 7.47, H$\gamma$/H$\beta$ = 0.40, & 5 & 14.5	& 14.0	& 6.0 &	37.9	& 11.7  \\
																				&& H$\delta$/H$\beta$ = 0.35 &&&&&\\
\noalign{\smallskip}\hline
\hline
\label{Cloudy-grid-new-resuts}
\end{tabular}
\begin{tablenotes}
$^1$Walter et al. 2012;  $^2$Tanaka et al. 2011, Das et al. 2015;  $^3$Anupama et al., 2013 ; $^4$Gherase et al. 2015, Burlak et al. 2015; $^5$Williams et al. 2017. 
\end{tablenotes}
\end{center}
\end{table*}

\section{Summary \& Discussions}
\label{section4}
We have computed grid models of novae  using photoionization code CLOUDY (Ferland et al. 2017).
The aim of this paper has been to generate an extended five-dimensional parameter space for dust-free
novae by varying hydrogen density, inner radius and thickness of the novae ejecta, temperature,
and luminosity of the ionizing source in commensurate with an observed range. From the model generated
synthetic spectra we have calculated line intensities of 56 hydrogen and helium lines which are generally observed 
prominently in novae spectra, spanning over a wide range of wavelengths: from ultraviolet to infrared.\\
\\
Simulated hydrogen and helium line intensities of our grid can be compared with
observations and physical parameters like $T_{BB}$, $L$, and $n_H$ can be inferred.
To test the robustness of our calculations, we have cross-checked predictions from our grid results for few novae e.g. RS Oph, V1065 Cen, V1974
Cyg, V1186 Sco and PW Vul with published data and they matched well. We also have estimated density and
luminosity of 5 other novae, one of which is an extragalactic nova. We have demonstrated that our grid models work fine. This gives us confidence that this will work successfully. However, there are few scopes to improve them and we discuss those in the following.\\
\\
We have considered here dust-free environment because from observations, it has been found that most of the novae do not form dust. 
Measurements of fluxes of spectral lines in post dust formation phase may not yield proper results. So, cautions should be taken while measuring the line fluxes. 
We plan to incorporate dust in our model and check the effect of dust on the line intensities.
Also, in the present calculation, we used solar abundance to limit the computational time. If abundances of other elements are increased or decreased, hydrogen line intensities may get changed.
To check this, we have constructed a few models with higher metallicity (2, 2.5 and 3 times solar metallicity) 
and found that variation in hydrogen line intensities are less than 15$\%$. 
We are in a process to extend our database with results for different metallicities in future.\\
\\
\begin{figure}
\centering
\includegraphics[width=3.4 in, height =3.1 in, clip]{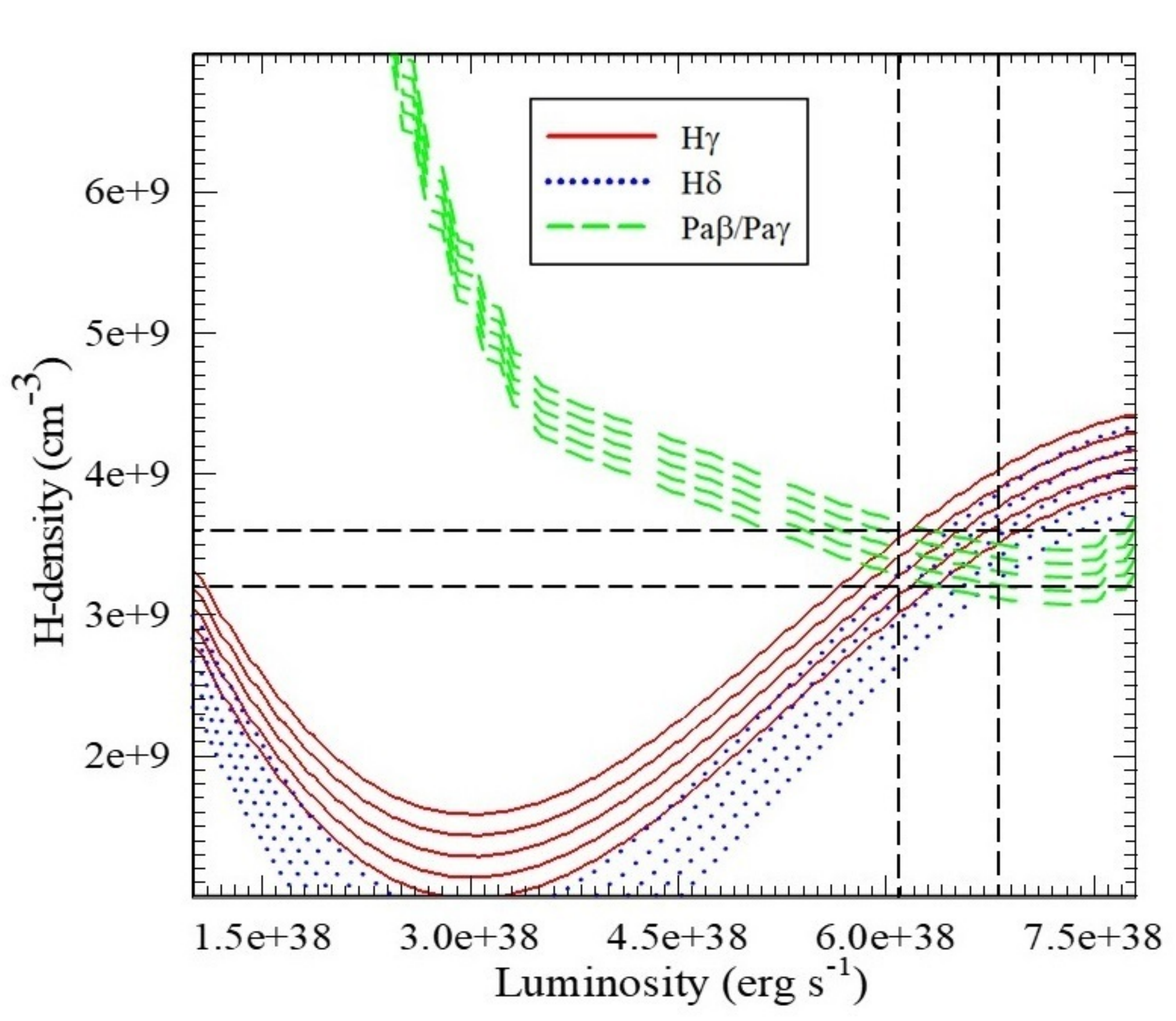}
\caption{Plot of different extracted contour lines of nova V5558 Sgr, 24 days after outburst, incorporating 5$\%$ error for observed flux values. The
red solid and blue dotted lines represent the extracted plots of line flux ratios of H$\gamma$ and H$\delta$ respectively w.r.t. H$\beta$ and the
green dashed line represents the extracted plots of line flux ratios of Pa$\beta$ w.r.t. Pa$\gamma$. The lines intersect between $L = (6.1 - 6.8) \times 10^{38}$ erg s$^{-1}$ and $n_H = (3.2 - 3.4) \times 10^9$ cm$^{-3}$. See section \ref{section4} for more details.}
\label{V5558Sgr-grid-error}
\end{figure}
In the present calculation, we have not incorporated errors associated with the measurements of the values of observables to keep our model simple. To check how the errors affect the results, we  have considered $\pm 5\%$ error associated with observed line fluxes for nova V5558 Sgr and shown the result in linear scale in Fig. 10. We have also checked in log scale that the curves do not intersect at any other points even after inclusion of the error margins. From Fig. 10, we find the modified results as $L = (6.1 - 6.8) \times 10^{38}$ erg s$^{-1}$ and $n_H = (3.2 - 3.4) \times 10^9$ cm$^{-3}$. These are in-line with the previous results of $L =  (6.4 \pm 0.4) \times 10^{38}  erg s^{-1}$ and $n_H = (3.4 \pm 0.2) \times 10^9 cm^{-3}$. Thus the uncertainties are less ( $\sim$ 6.25$\%$ in case of $L$ and $\sim$ 5.8$\%$ in case of $n_H$), which ensures the applicability of the method.
If uncertainties in the observables increase, the results will also become more uncertain.
So, cautions should be taken while choosing the emission lines. It is suggested to consider strong emission lines (e.g. H$\alpha$, H$\beta$, H$\gamma$, Pa$\alpha$, Pa$\beta$, Pa$\gamma$, Br$\gamma$ etc.) to reduce the error during measurements.\\
\\
The nature of the contours depend on the line ratios, and line intensity vary from novae to novae.
In few cases, degeneracy may appear i.e. more than one countours may follow the same locus or have multiple intersections.  For example in Fig. 6, the curves nearly intersect around log(L) = 38.2 [in erg s$^{-1}$] and log(n$_H$) = 11.5 [in cm$^{-3}$] in addition to intersection at log(L) = 38.8 [in erg s$^{-1}$] and log(n$_H$) = 9.5 [in cm$^{-3}$]. In such cases, results obtained from other calculations may help to estimate the parameter values. Also, more emission lines/observables available, may be used to ensure that we do not get any other intersections. It is suggested to consider strong emission lines (H$\alpha$, H$\beta$, H$\gamma$, Pa$\alpha$, Pa$\beta$, Pa$\gamma$, Br$\gamma$ etc.) with higher flux ratios as they include less instrumentation errors during measurements. Generally the degeneracy do not increase while considering more observables. In such case, we should take emission lines of other elements (e.g. He line for nova V1186 Sco) also.\\
\\
If we take the line ratios w.r.t. other lines instead of H$\beta$, shapes of the contours will be different and degeneracy may not occur. As we understand from the plots, strong lines like the lower members of the hydrogen series have a less tendency to be degenerate. So, multi-wavelength observations would be useful in such cases.
If there are multiple intersections, results obtained from other calculations may help to estimate the parameter values.
We should use as many emission lines/observables available to make sure that multiple intersections do not occur.  If no intersection is found, a range of the values could be estimated.\\
\\
All of our models presented here are one-dimensional and radiation from central source has been
approximated by a black body radiation. It would be interesting
to study how the advanced stellar atmospheres instead of the
simple black body radiation could affect results.
In future, we would like to include wind in our models and extend it to 3-D using an advanced version of pyCloudy.
Further more, a finer mesh can be utilized to determine more precise values of the physical parameters.
In addition to optical and infrared, X-ray spectra have been observed in many novae (e.g. V2491 Cyg, Ness et al. 2011; V4743 Sgr, Ness et al. 2003 etc.) and similar grid models can also be computed for X-ray spectra. CLOUDY can be used to model all these lines.\\
\\
For the benefit of all the astronomical community, we have kept the database online at
{\it https://aninditamondal1001.wixsite.com/researchworks} under ``\textsc{Novae Grid Data}" on the page {\it Research/List-of-Documents}. Also a {\it ``Read Me"} file has been kept which describes the structure of the datafiles. Data can be obtained individually
in *.xls format (16.4 KB each). The whole grid (256 tables) has a total size of 4.1 MB.\\
\\
\section{Acknowledgement}
The research work at S N Bose National Centre for Basic Sciences is funded by the Department of Science and Technology, Government of India. GS would like to acknowledge Department of Science and Technology, Government of India, for her WOS-A project SR/WOS-A/PM-9/2017. The authors also thank the anonymous referee for valuable comments and suggestions that have helped to improve the quality of the manuscript.

\end{document}